\documentclass[preprint,12pt]{elsarticle}
\usepackage{amssymb}
\usepackage{lineno}
\journal{}

\begin{document}

\begin{frontmatter}

\title{SiPM module for the ACME III electron EDM search}

\author[okayama]{A. Hiramoto}
\author[okayama]{T. Masuda}
\author[harvard]{D. G. Ang}
\author[harvard]{C. Meisenhelder}
\author[ucb]{C. Panda}
\author[okayama]{N. Sasao}
\author[okayama]{S. Uetake}
\author[harvard,chicago]{X. Wu}
\author[chicago]{D. Demille}
\author[harvard]{J. M. Doyle}
\author[northwestern]{G. Gabrielse}
\author[okayama]{K. Yoshimura}

\affiliation[okayama]{organization={Research Institute for Interdisciplinary Science, Okayama University},%Department and Organization
            city={Okayama},
            postcode={700-8530}, 
            country={Japan}}

\affiliation[harvard]{organization={Department of Physics, Harvard University},%Department and Organization
            addressline={Address Two}, 
            city={Cambridge},
            postcode={02138}, 
            state={MA},
            country={USA}}

\affiliation[chicago]{organization={James Franck Institute and Department of Physics, University of Chicago},%Department and Organization
            city={Chicago},
            postcode={60637}, 
            state={IL},
            country={USA}}

\affiliation[northwestern]{organization={Center for Fundamental Physics, Northwestern University},%Department and Organization
            city={Evanston},
            postcode={60208}, 
            state={IL},
            country={USA}}

\affiliation[ucb]{organization={Department of Physics, University of California},%Department and Organization
            city={Berkeley},
            postcode={94720}, 
            state={CA},
            country={USA}}

\begin{abstract}

This report shows the design and the performance of a large area Silicon Photomultiplier (SiPM) module developed detection of fluorescent light emitted from a $\sim$10cm scale volume. The module was optimized for the planned ACME III electron electric dipole moment (eEDM) search, which will be a powerful probe for the existence of physics beyond the Standard Model of particle physics. The ACME experiment searched for the eEDM with the world’s highest sensitivity using cold ThO polar molecules (ACME II\cite{acme2018}). In ACME III, SiPMs will be used for detection of fluorescent photons (the fundamental signal of the experiment) instead of PMTs, which were used in the previous measurement. We have developed an optimized SiPM module, based on a 16-channel SiPM array. Key operational parameters are charac- terized, including gain and noise. The SiPM dark count rate, background light sensitivity, and optical crosstalk are found to all be well suppressed and more than sufficient for the ACME III application.

\end{abstract}


\end{frontmatter}

%\linenumbers

\section{Introduction}
\label{sec:introduction}

The electron electric dipole moment (eEDM) is a powerful probe for physics beyond the Standard Model. The Standard Model of particle physics does not have enough CP violation to explain the matter-antimatter asymmetry of the universe. This has driven the creation of many experiments to look for a new source of CP violation. An EDM measurement is one of the CP-violation searches because the existence of EDM corresponds to a violation of T-symmetry, which is equal to CP violation under the assumption of the CPT conservation. 

\begin{figure}[h]
\centering
\includegraphics[width=12cm]{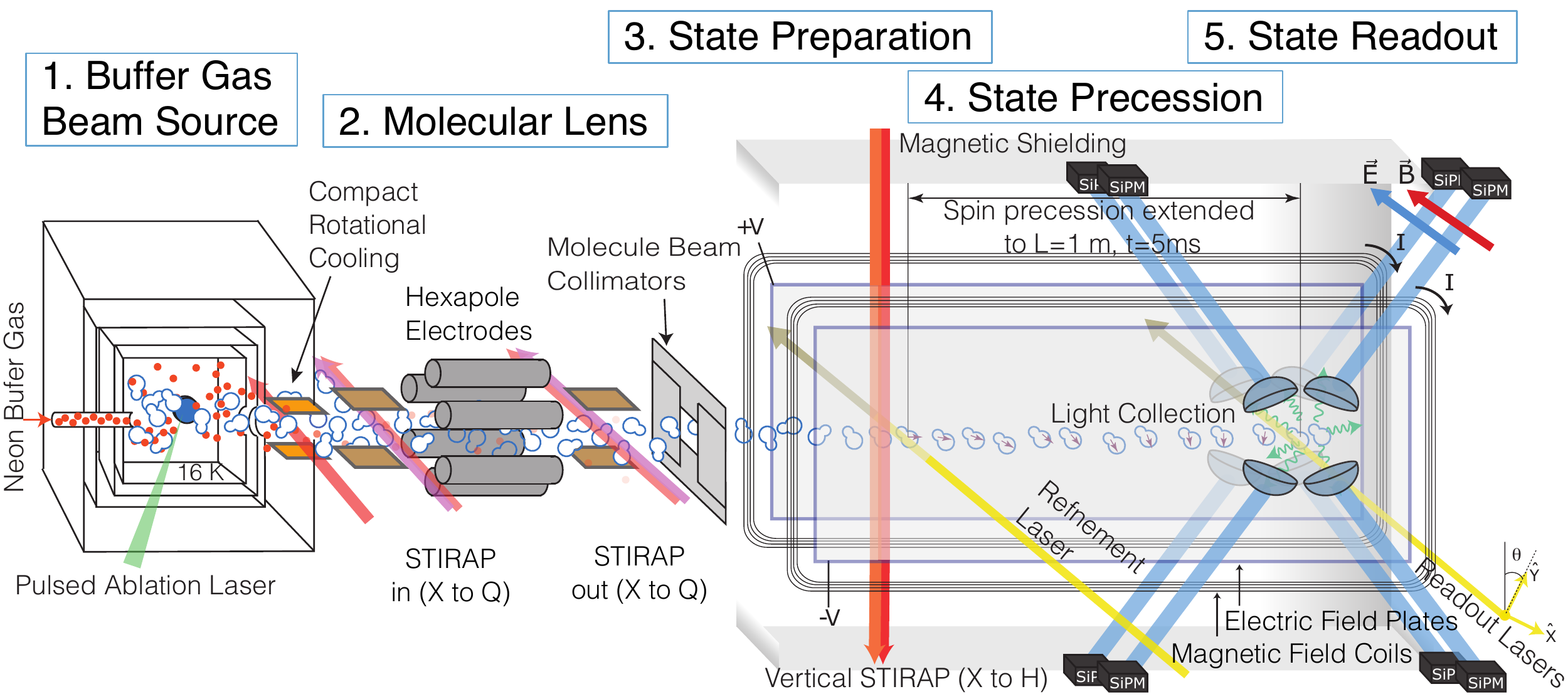}
\caption{\label{fig:acme} Schematic view of the ACME experiment. Fluorescence photons from ThO molecules are detected by SiPMs.}
\end{figure}

The ACME experiment has twice set the most stringent limits on eEDM\cite{acme2014,acme2018}. Our approach is to use a cold beam of thorium monoxide (ThO) molecules in a spin precession setup. ThO has a very large effective electric field\cite{skripnikov2016,denis2016}, which would give rise to a large electron spin precession due to an eEDM. More detail about the ACME measurement can be found in\cite{acme2018,acme2014}. In the next generation of the ACME experiment (ACME III), we plan to improve the sensitivity about 30 times through upgrades to the apparatus. Figure \ref{fig:acme} shows the apparatus for ACME III. A cryogenic buffer-gas beam source generates a cold ThO beam, and hexapole electrodes focus that beam into the interaction region increasing the number of detected molecules\cite{wu2022}. At the entrance of the spin precession region, molecules are optically driven into the H state for the eEDM measurement, where the unpaired electron spins precess in combined electric and magnetic fields. At the end of the precession region, the molecules in the H state are excited to the I state using laser excitation. The excited molecules emit 512\,nm fluorescence light, which characterizes the phase of the spin precession, the key parameter in the eEDM measurement. The peak flux of detected photons at each photodetector is approximately 1000\,photoelectrons/$\mu$s.

In the previous measurements, ACME used PMTs for signal readout. We plan to replace the PMTs with Silicon Photomultipliers (SiPMs) to increase the photon detection efficiency. In Sec. \ref{sec:design}, we introduce the SiPM module, which can be used as a simple replacement for a PMT. In Sec. \ref{sec:characterization}, several important characteristics of the module are described. In Sec. \ref{sec:mount}, mounting of the SiPM module to the ACME apparatus is introduced, and a test using the ThO beam is described in Sec. \ref{sec:tho}. Finally, we summarize the characteristics of the SiPM module in Sec. \ref{sec:summary}.

\section{Design}
\label{sec:design}
We will employ a 16-channel SiPM array S13361-6075NE-04 (Hamamatsu Photonics K.K.) for ACME III. Table \ref{table:comparison} shows comparisons between ACME II PMT and ACME III SiPM. Replacing the PMTs with the SiPMs leads to a quantum efficiency (Q.E.) gain at 512\,nm from around 25\% to 45\%. On the other hand, detection efficiency at 703\,nm is also increased, which is potentially problematic as 703\,nm light is used to excite the molecules. Thus, the 703\,nm light can potentially lead to increased background noise. Another potential drawback is the excess noise factor of the SiPM is as large as that of the PMT due to large optical crosstalk. Finally, the SiPMs have much higher dark count rate (DCR) than the PMTs at the same ambient temperature, and the SiPMs also have relatively slow response time due to their large capacitance.
Taking into consideration the issues mentioned above, there were several design goals for our SiPM module: suppressing background at 703\,nm to a level at least as low as the PMT, achieve a DCR lower than 80\,kcps/ch, having sufficient optical crosstalk suppression, and achieving a response faster than the I state lifetime (115\,ns). We also aimed for good output linearity, up to 1000\,photoelectrons/$\mu$s. 

\begin{table}[h]
 \caption{Comparison between ACME II PMT and ACME III SiPM. For the SiPM, catalog specks before constructed as a module are shown.}
 \label{table:comparison}
 \centering
  \begin{tabular}{lcc}
   \hline
    & ACME II PMT & ACME III SiPM \\
    \hline \hline
    Model number & R7600U-300 & S13361-6075NE-04\\
    Sensitive area & 18\,mm$\times$18\,mm & 24\,mm$\times$24\,mm (16 channels)\\
    Q.E. at 512\,nm & $\sim$25\% & $\sim$45\%\\
    Q.E. at 703\,nm & $\sim$0.6\% & $\sim$20\%\\
    Excess noise factor & $\sim$1.2 & $\sim$1.2\\
    DCR at room temperature & $\sim$3\,kcps & $\sim$2\,Mcps/channel\\
    Capacitance & few pF & 1.4\,nF\\
   \hline
  \end{tabular}
\end{table}

\begin{figure}[h]
\centering
\begin{minipage}{0.32\textwidth}
\centering
\includegraphics[height=5.3cm]{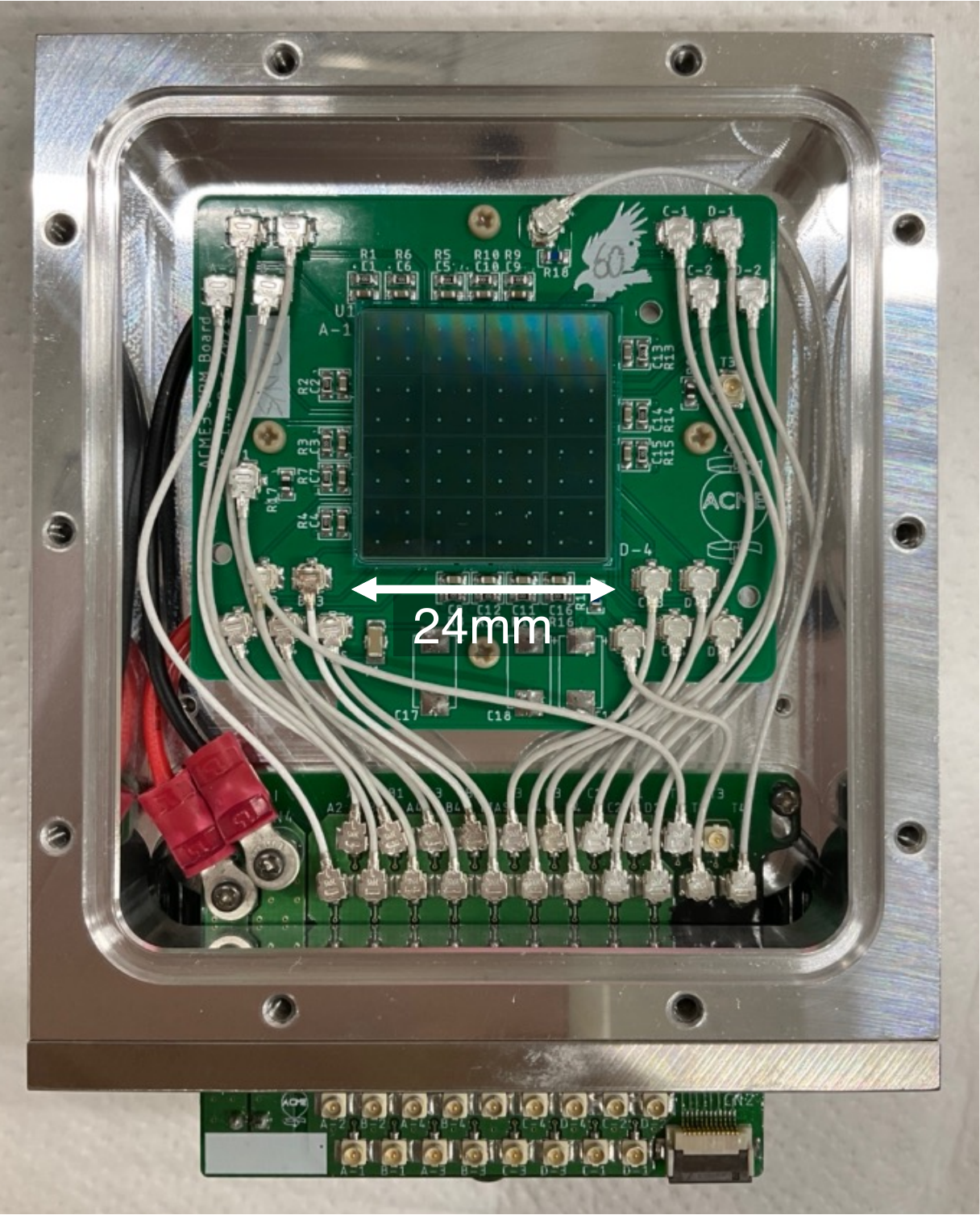}
\end{minipage}
\begin{minipage}{0.62\textwidth}
\centering
\includegraphics[height=5.3cm]{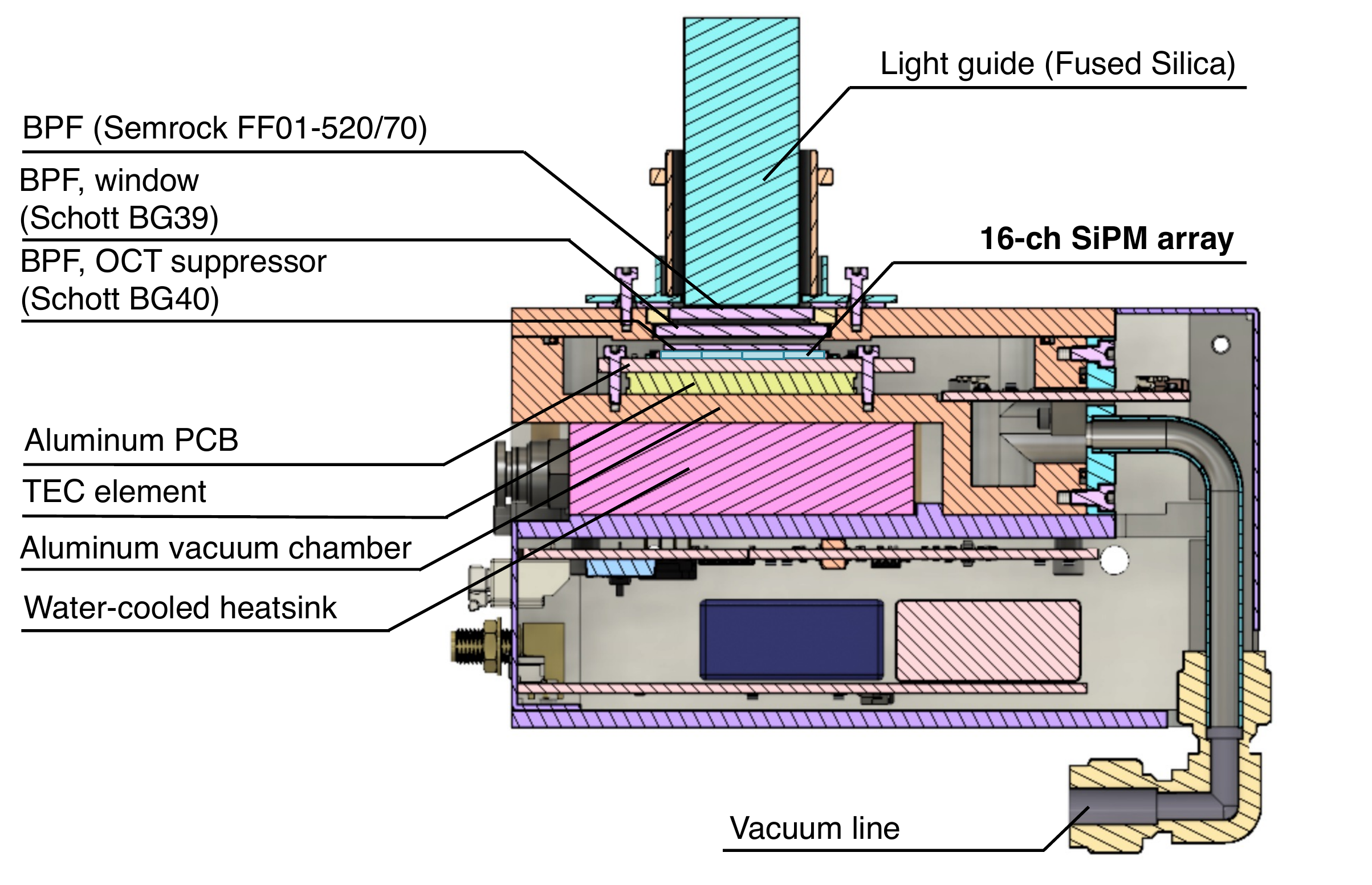}
\end{minipage}
\caption{\label{fig:sipm}Photo (left) and a cross-sectional view (right) of the SiPM module.}
\end{figure}

Figure \ref{fig:sipm} shows a photo and a schematic of the SiPM module. The DCR, the most serious concern, is suppressed by cooling the SiPM array using a thermoelectric cooler (TEC, Thorlabs Inc. TECH4). The SiPM array is mounted in vacuum on a dedicated printed circuit board (PCB) made of aluminum for better thermal conductivity. A water-cooled heatsink is placed on the backside of the vacuum chamber and the water temperature is controlled to be 20$^\circ$C. Three wavelength-selective optical filters are placed above the SiPM array to reduce optical cross talk\cite{masuda2021} and suppress the 703\,nm background.

On the backside of the chamber, PCBs for SiPM bias and readout circuits are placed. We apply bias voltage without a bias adjustment system for each channel in a SiPM array because the breakdown voltage between the 16 channels are uniform enough. The output signal of the 16 channels are summed, and a pole-zero cancellation technique\cite{gola2013} realizes a fast signal response. A three-pole Bessel filter shapes the summed signal to improve the signal-to-noise ratio and reduce the digitizing sampling rate. The single photoelectron signal has about peak voltage of 2-mV and 80-ns full width at half maximum when operated with 3.0\,V overvoltage.

\section{SiPM module performance}
\label{sec:characterization}

\begin{figure}[h]
\centering
\begin{minipage}{0.47\textwidth}
\centering
\includegraphics[height=5.0cm]{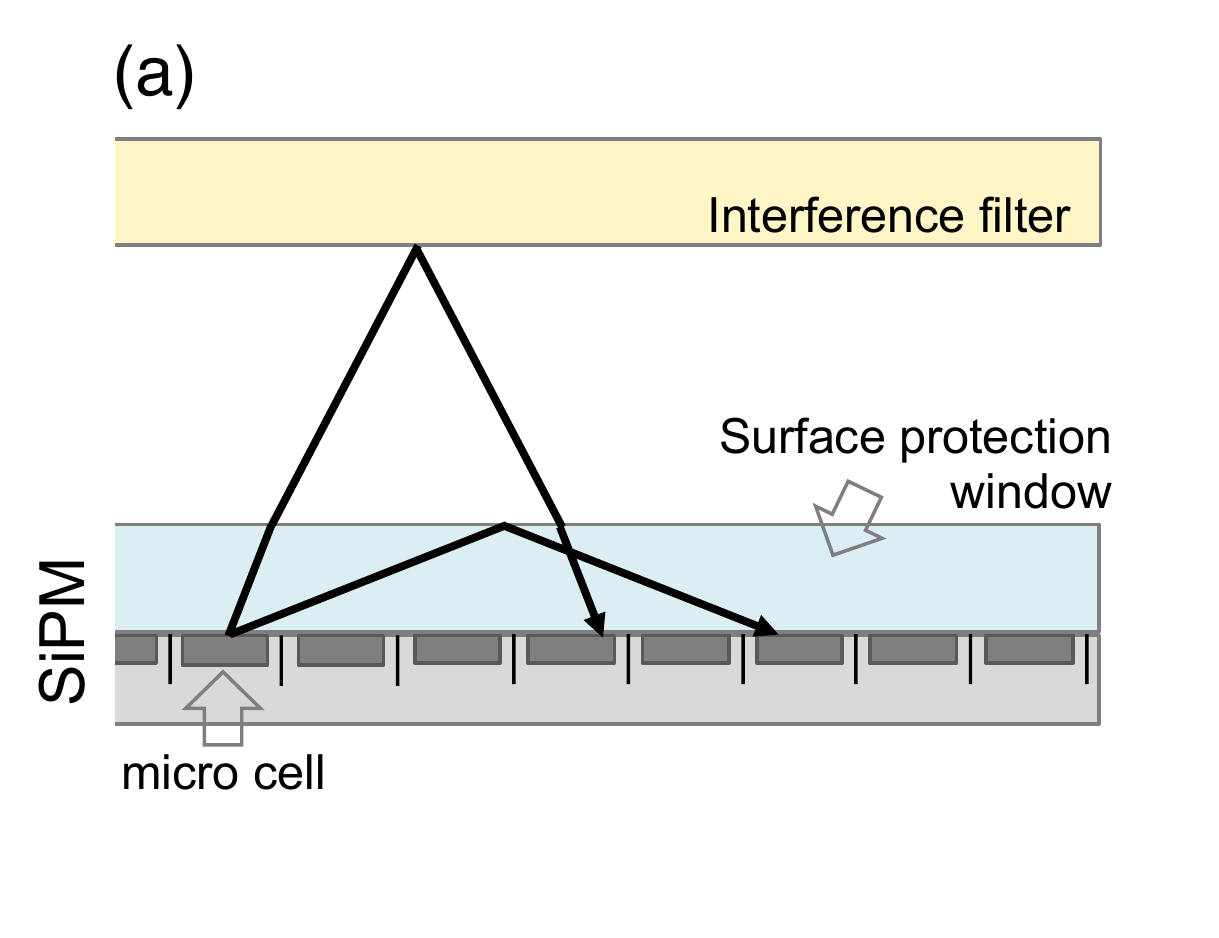}
\end{minipage}
\begin{minipage}{0.47\textwidth}
\centering
\includegraphics[height=5.0cm]{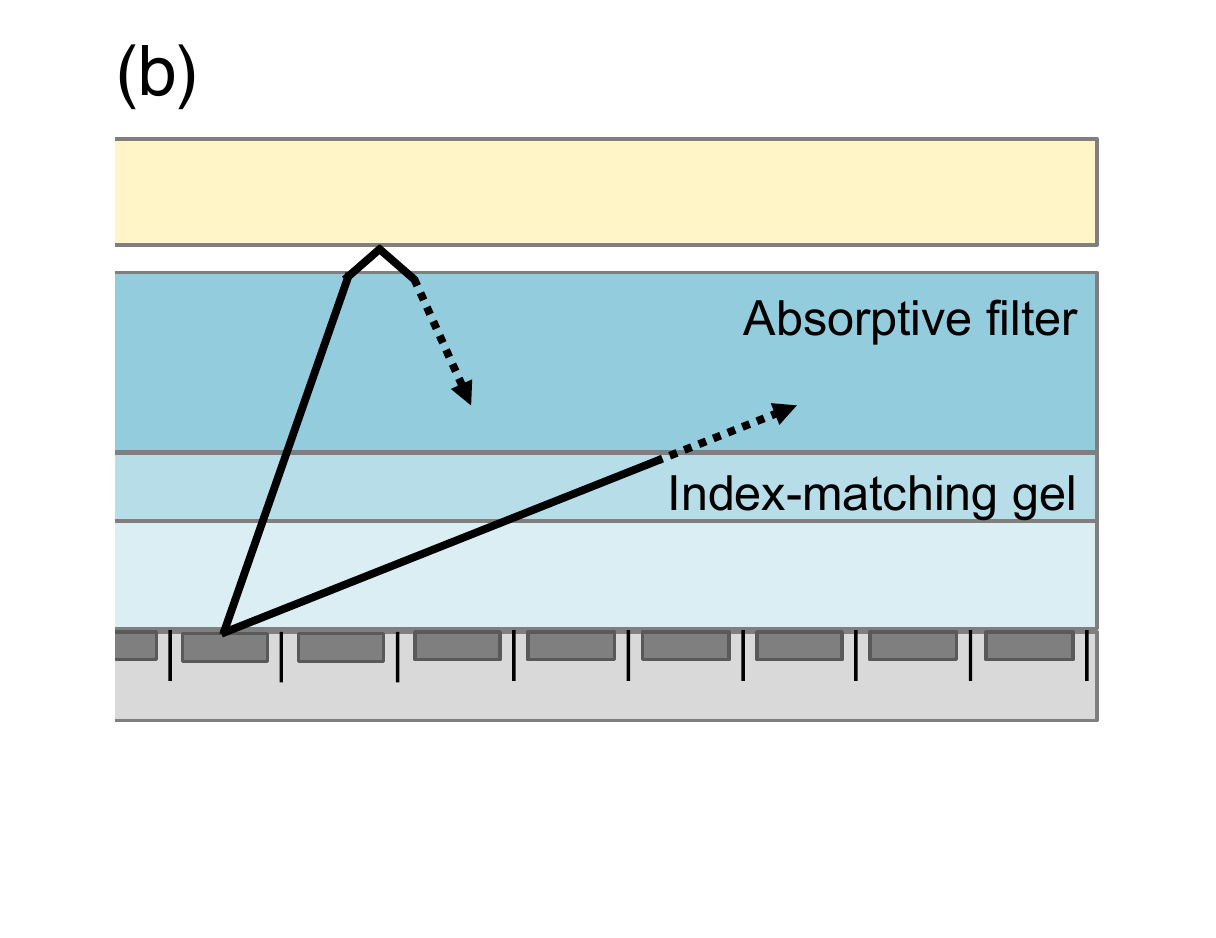}
\end{minipage}
\caption{\label{fig:oct}a) Mechanism of the OCT. An interference filter is placed above the SiPM. b) OCT suppression mechanism using an absorptive filter.}
\end{figure}

In this section, we describe the characterization of the SiPM module, especially the optical cross talk (OCT) and the excess noise factor. The photon counting resolution of the SiPM module is important for eEDM sensitivity. However, crosstalk and afterpulsing can degrade the photon-counting resolution. As we mentioned in Sec. \ref{sec:design}, one of the three wavelength-selective optical filters is placed for the OCT suppression. The mechanism for the OCT is shown in Fig. \ref{fig:oct}(a). Secondary photons produced during the avalanche process can be reflected from the surface of the SiPM coating or from the filters above it, and then detected by other SiPM micro-cells. Since we use an array type SiPM, the OCT occurs both in the same channel and inter-channels. To suppress the OCT, we placed an absorptive filter (SCHOTT BG40) directly above the SiPM using index-matching gel, see Fig. \ref{fig:oct}(b). This filter has a transmission peak at 500\,nm. The filter absorbs photons along the paths shown in Fig. \ref{fig:oct}(b), and the OCT probability is reduced from 24.7\% to 4.3\%, compared to the case where only the interference filter (Semrock FF01-520/70) is used, as shown in Fig. \ref{fig:oct}(a). More detail about this study can be found in reference \cite{masuda2021}.

\begin{figure}[h]
\centering
\includegraphics[height=6.0cm]{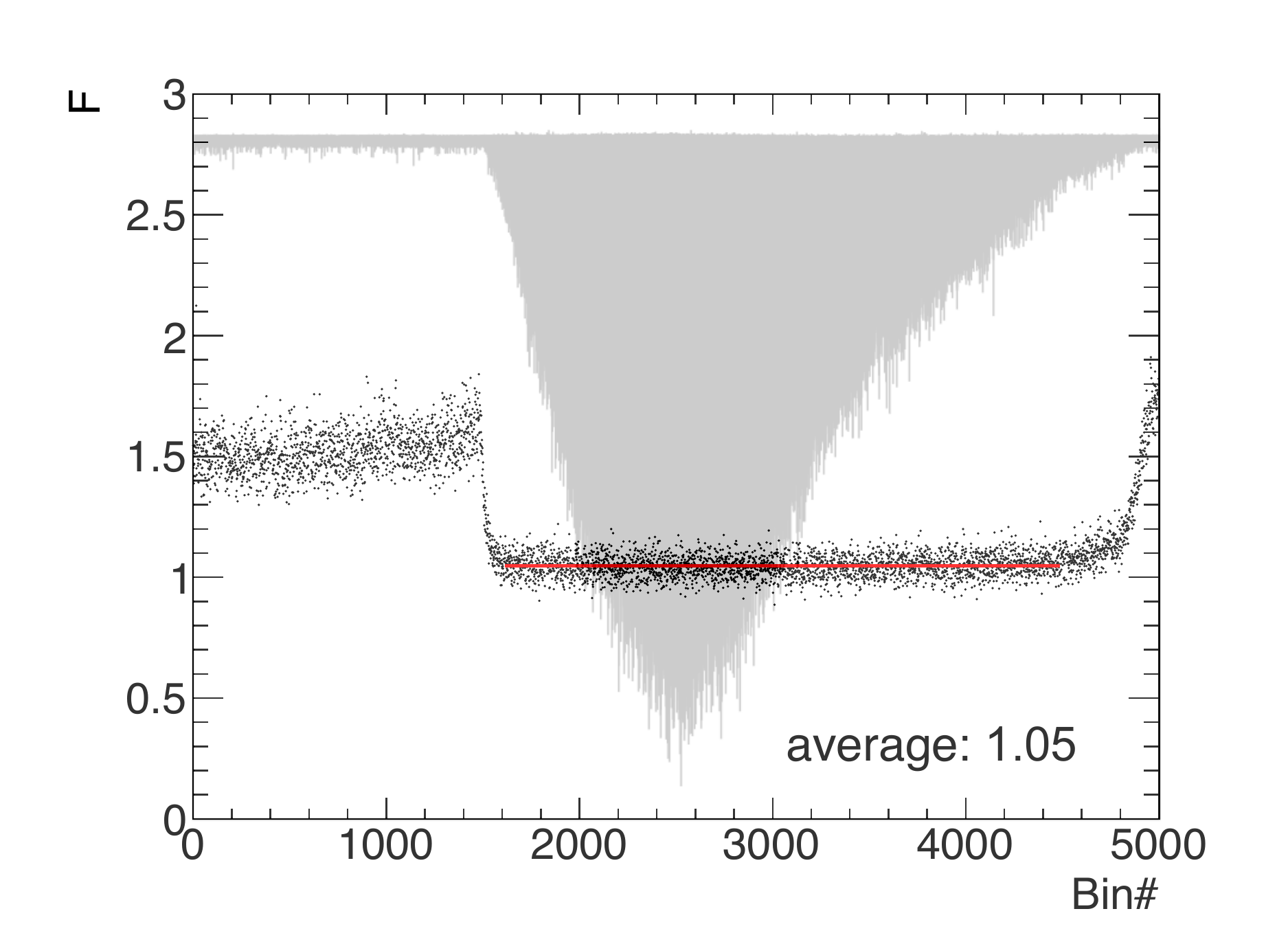}
\caption{\label{fig:pulse} Example of the Fano factor at each bin. Overall envelope of the square pulses created by a 515\,nm laser diode is shown in the gray plot. The Fano factor
in the bins without light injection are not estimated properly due to fluctuation of the signal baseline.}
\end{figure}

An effect of the OCT and afterpulsing is characterized by a Fano factor ($F$) \cite{Vinogradov2009}. When pulsed photons are injected using a light source, the value $F$ can be calculated from the standard deviation ($\sigma$) and the mean value ($\mu$) of the observed number of photons as $\sigma^2/\mu$\cite{teich1986}. Here, note that we estimate the number of photons using the effective gain, which is calculated by multiplying the single-photoelectron gain by 1.1. The factor 1.1 comes from our previous measurement which suggests that the OCT and afterpulsing probability is around 10\%. A 515\,nm laser diode (OSRAM PLT5) was used with an arbitrary waveform generator (siglent SDG2042X), and the SiPM output waveforms were recorded by a 2-channel oscilloscope (NI PXIe-5164). The SiPM module was cooled down to -15$^\circ$C by the TEC. The gray plot in Fig. \ref{fig:pulse} shows the trace of the injected signal envelope, which reproduces the expected ACME molecular beam signal. The trace has 400\,kHz square pulse structure, which corresponds to the laser polarization switching which is used in ACME. Each pulse has about 1\,$\mu$s width, and we call each pulse a "bin". At the peak of the trace in Fig. \ref{fig:pulse}, a bin has around 1000 detected photons. The number of photons in each bin is measured, and the $\sigma$ and $\mu$ of the detected number of photons are obtained for each bin from 1000-shot waveforms.

An example of the Fano factor at each bin is shown in Fig. \ref{fig:pulse}. We have 10 SiPM modules, and we find that the average of $F$ is around 1.07. This result suggests that the excess noise by the OCT and afterpulsing is reduced to around 7\% owing to the OCT suppression.

We also carried out several performance tests with 10 SiPM modules. The single-photoelectron gain is so uniform that the effect on photon counting resolution is negligible compared to the excess noise measured above. The DCR at -15$^\circ$C is suppressed to 3\% of that at room temperature and meets the ACME requirement (80\,kcps/channel) well. In addition, we carried out a long-term stability check. After a 7-weeks of continuous operation (corresponding to 1.7$\times10^{14}$ detected photons), we found that the fluctuation of the single photon gain and the photon detection efficiency was less than $\pm$1\%. A paper describing these characterizations is in preparation\cite{masuda2022}.

\section{Mounting on the ACME apparatus}
\label{sec:mount}

\begin{figure}[h]
\centering
\includegraphics[width=5.5cm]{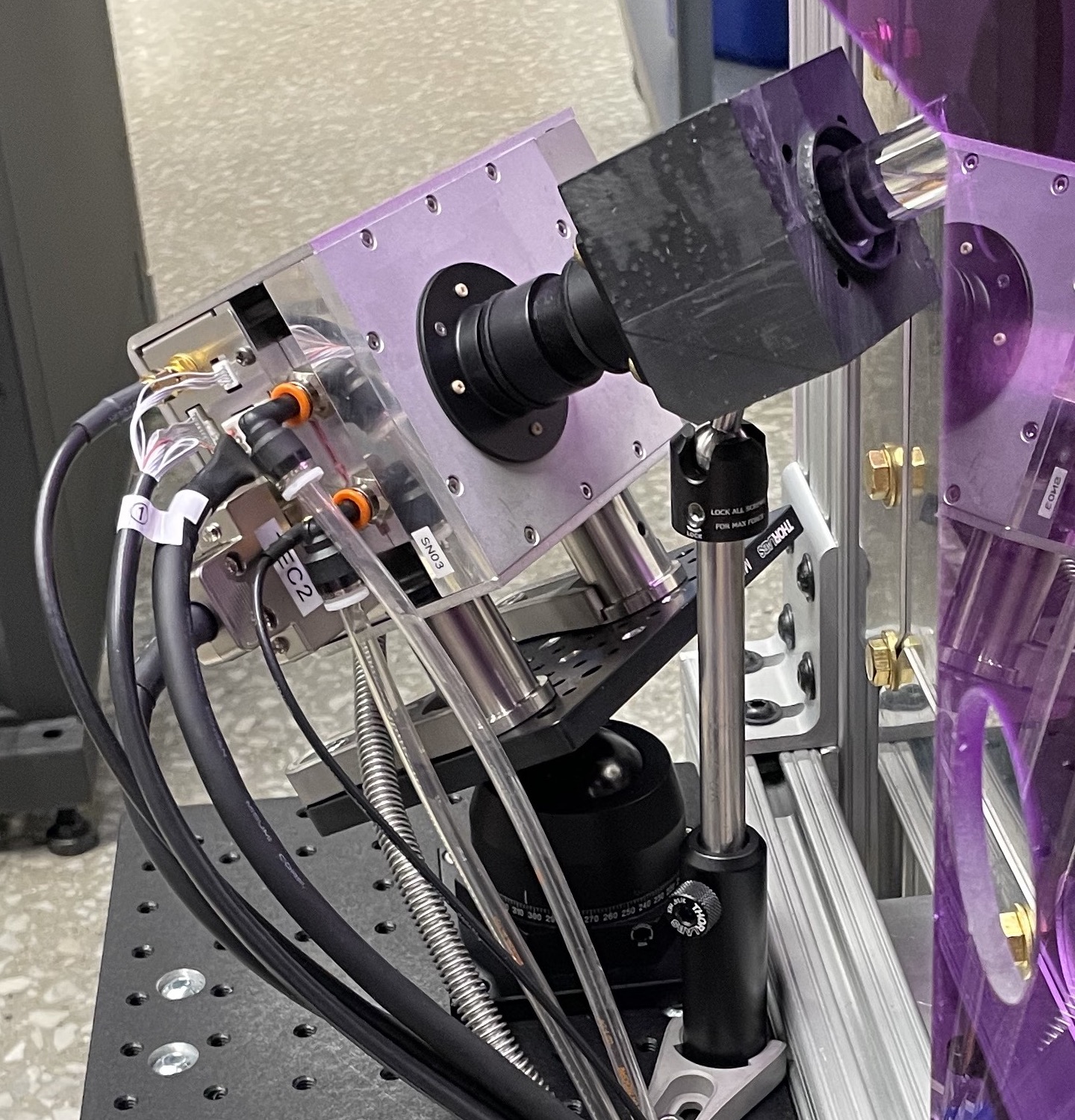}
\caption{\label{fig:mounting}Photo of the SiPM module mounting. The light pipe sleeve and the rubber bellows are not attached yet.}
\end{figure}

The 512\,nm photons in the ACME measurement are transmitted to the SiPM modules by light collecting lens and light pipes. Figure \ref{fig:mounting} shows a photo of the mounting of the light pipe and the SiPM module. The SiPM module is mounted on a breadboard with an articulating base stage (THORLABS SL20) so that the surface of the SiPM module can be set parallel to the light pipe surface. The light pipe is held by a mounting structure as shown in Fig. \ref{fig:mounting}. In order to suppress room light, a sleeve is attached to the light pipe for shading, and a rubber bellows covers the section between the light pipe mounting and the SiPM module surface. With these structures, we achieved the photon counting rate consistent with the DCR, even with the room light is turned on.

\section{Measurement with ACME ThO beam}
\label{sec:tho}

\begin{figure}[h]
\centering
\includegraphics[width=7cm]{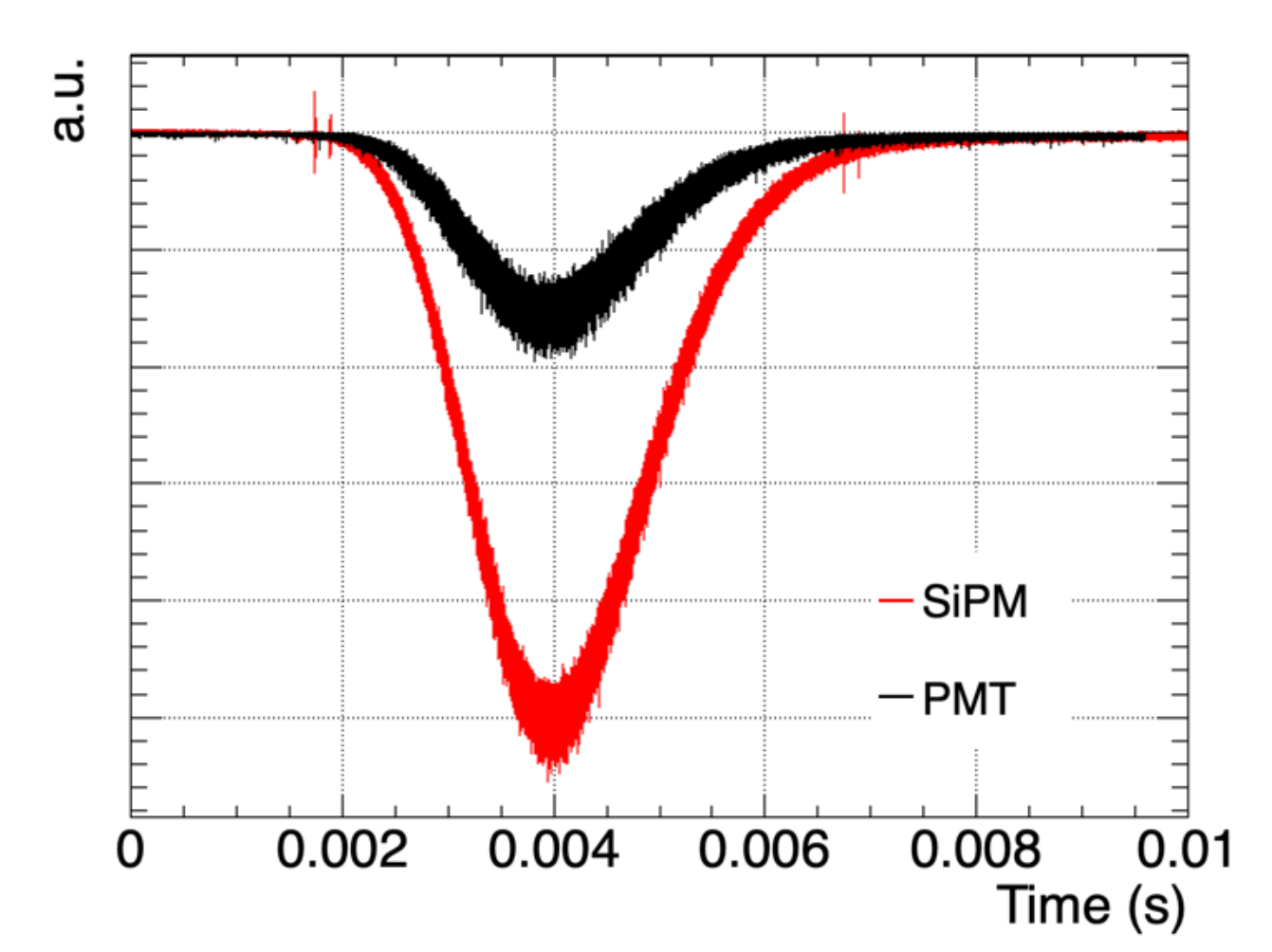}
\caption{\label{fig:result}Comparison of the signal yields between the PMT and the SiPM with normalization of single photoelectron gain and beam intensity monitor.}
\end{figure}

We carried out a measurement using a ThO beam to compare photon detection rates between the ACME II PMTs and the ACME III SiPM modules. Details of this measurement are also described in \cite{masuda2022}. The cold ThO molecules from the beam source are focused using a molecular lens and excited at the detection region using a laser light. The 512\,nm spontaneously emitted photons are detected by the SiPM module or PMT through the same light-collection optics (lens and light pipes). We placed a PMT as a reference to monitor the shot-by-shot variation of ThO beam intensity. Figure \ref{fig:result} shows the signal yields after normalizing by single photoelectron gains and the beam intensity monitor. The SiPM module has about three times higher detection rate than the PMT, including the improvement in both quantum efficiency and geometrical acceptance of the SiPM mpdule.

\section{Summary}
\label{sec:summary}
The ACME III plans on using SiPMs as photon detectors, instead of the PMTs. We developed the dedicated SiPM module and characterized and optimized its performance. The optical crosstalk and the dark count rate are well suppressed, and the performance of the SiPM module meets the ACME III requirements, with performance superior to the previously used PMTs. The modules were compared with PMTs in a realistic beam setup (the ACME II apparatus), and data indicate that the SiPM modules have about three times higher detection efficiency than PMTs.

\section*{Acknowledgement}
This work is supported by Okayama University (RECTOR program), Japan Society for the Promotion of Science (KAKENHI JP20KK0068, JP21H01113, JP21J01252), Matsuo Foundation, National Science Foundation, Gordon and Betty Moore Foundation, and Alfred P. Sloan Foundation.

 \bibliographystyle{elsarticle-num} 
 \bibliography{cas-refs}

\begin{thebibliography}{10}
\expandafter\ifx\csname url\endcsname\relax
  \def\url#1{\texttt{#1}}\fi
\expandafter\ifx\csname urlprefix\endcsname\relax\def\urlprefix{URL }\fi
\expandafter\ifx\csname href\endcsname\relax
  \def\href#1#2{#2} \def\path#1{#1}\fi

\bibitem{acme2018}
V.~Andreev, D.~G. Ang, D.~DeMille, J.~M. Doyle, G.~Gabrielse, J.~Haefner, N.~R.
  Hutzler, Z.~Lasner, C.~Meisenhelder, B.~R. O'Leary, C.~D. Panda, A.~D. West,
  E.~P. West, X.~Wu, {Improved limit on the electric dipole moment of the
  electron}, Nature 562~(7727) (2018) 355--360.
\newblock \href {http://dx.doi.org/10.1038/s41586-018-0599-8}
  {\path{doi:10.1038/s41586-018-0599-8}}.

\bibitem{acme2014}
J.~Baron, W.~C. Campbell, D.~DeMille, J.~M. Doyle, G.~Gabrielse, Y.~V.
  Gurevich, P.~W. Hess, N.~R. Hutzler, E.~Kirilov, I.~Kozyryev, B.~R. O'Leary,
  C.~D. Panda, M.~F. Parsons, E.~S. Petrik, B.~Spaun, A.~C. Vutha, A.~D. West,
  Order of magnitude smaller limit on the electric dipole moment of the
  electron, Science 343~(6168) (2014) 269--272.
\newblock \href {http://dx.doi.org/10.1126/science.1248213}
  {\path{doi:10.1126/science.1248213}}.

\bibitem{skripnikov2016}
L.~V. Skripnikov, Combined 4-component and relativistic pseudopotential study
  of {ThO} for the electron electric dipole moment search, The Journal of
  Chemical Physics 145~(21) (2016) 214301.
\newblock \href {http://dx.doi.org/10.1063/1.4968229}
  {\path{doi:10.1063/1.4968229}}.

\bibitem{denis2016}
M.~Denis, T.~Fleig, In search of discrete symmetry violations beyond the
  standard model: Thorium monoxide reloaded, The Journal of Chemical Physics
  145~(21) (2016) 214307.
\newblock \href {http://dx.doi.org/10.1063/1.4968597}
  {\path{doi:10.1063/1.4968597}}.

\bibitem{wu2022}
X.~Wu, P.~Hu, Z.~Han, D.~G. Ang, C.~Meisenhelder, G.~Gabrielse, J.~M. Doyle,
  D.~DeMille,
  \href{https://iopscience.iop.org/article/10.1088/1367-2630/ac8014}{{Electrostatic
  focusing of cold and heavy molecules for the ACME electron EDM search}}, New
  Journal of Physics 24~(7) (2022) 073043.
\newblock \href {http://arxiv.org/abs/2204.05906} {\path{arXiv:2204.05906}},
  \href {http://dx.doi.org/10.1088/1367-2630/ac8014}
  {\path{doi:10.1088/1367-2630/ac8014}}.
\newline\urlprefix\url{https://iopscience.iop.org/article/10.1088/1367-2630/ac8014}

\bibitem{masuda2021}
T.~Masuda, D.~G. Ang, N.~R. Hutzler, C.~Meisenhelder, N.~Sasao, S.~Uetake,
  X.~Wu, D.~DeMille, G.~Gabrielse, J.~M. Doyle, K.~Yoshimura, {Suppression of
  the optical crosstalk in a multi-channel silicon photomultiplier array},
  Optics Express 29~(11) (2021) 16914.
\newblock \href {http://arxiv.org/abs/2105.01519} {\path{arXiv:2105.01519}},
  \href {http://dx.doi.org/10.1364/OE.424460} {\path{doi:10.1364/OE.424460}}.

\bibitem{gola2013}
A.~Gola, C.~Piemonte, A.~Tarolli, {Analog Circuit for Timing Measurements With
  Large Area SiPMs Coupled to LYSO Crystals}, IEEE Transactions on Nuclear
  Science 60~(2) (2013) 1296--1302.
\newblock \href {http://dx.doi.org/10.1109/TNS.2013.2252196}
  {\path{doi:10.1109/TNS.2013.2252196}}.

\bibitem{Vinogradov2009}
S.~Vinogradov, T.~Vinogradova, V.~Shubin, D.~Shushakov, K.~Sitarsky,
  \href{https://doi.org/10.1109/nssmic.2009.5402300}{Probability distribution
  and noise factor of solid state photomultiplier signals with cross-talk and
  afterpulsing}, in: 2009 {IEEE} Nuclear Science Symposium Conference Record
  ({NSS}/{MIC}), {IEEE}, 2009.
\newblock \href {http://dx.doi.org/10.1109/nssmic.2009.5402300}
  {\path{doi:10.1109/nssmic.2009.5402300}}.
\newline\urlprefix\url{https://doi.org/10.1109/nssmic.2009.5402300}

\bibitem{teich1986}
M.~Teich, K.~Matsuo, B.~Saleh,
  \href{https://doi.org/10.1109/jqe.1986.1073137}{Excess noise factors for
  conventional and superlattice avalanche photodiodes and photomultiplier
  tubes}, {IEEE} Journal of Quantum Electronics 22~(8) (1986) 1184--1193.
\newblock \href {http://dx.doi.org/10.1109/jqe.1986.1073137}
  {\path{doi:10.1109/jqe.1986.1073137}}.
\newline\urlprefix\url{https://doi.org/10.1109/jqe.1986.1073137}

\bibitem{masuda2022}
T.~Masuda, H.~Ayami, D.~G. Ang, C.~Meisenhelder, P.~D. Cristian, N.~Sasao,
  S.~Uetake, X.~Wu, D.~DeMille, G.~Gabrielse, J.~M. Doyle, K.~Yoshimura,
  {High-sensitivity low-noise photodetector using large-area silicon
  photomultiplier}.

\end{thebibliography}





% \end{thebibliography}
\end{document}